\documentclass{article}

\usepackage{PRIMEarxiv}

\usepackage[utf8]{inputenc} % allow utf-8 input
\usepackage[T1]{fontenc}    % use 8-bit T1 fonts
\usepackage{hyperref}       % hyperlinks
\usepackage{url}            % simple URL typesetting
\usepackage{booktabs}       % professional-quality tables
\usepackage{amsfonts}       % blackboard math symbols
\usepackage{nicefrac}       % compact symbols for 1/2, etc.
\usepackage{microtype}      % microtypography
\usepackage{lipsum}
\usepackage{fancyhdr}       % header
\usepackage{graphicx}       % graphics
\graphicspath{{media/}}     % organize your images and other figures under media/ folder
\usepackage{pdflscape}

\usepackage{xcolor}
\title{Automating Attack Graph Construction for Agentic Pentesting. Towards Neuro-Symbolic Vulnerability Hunting.\thanks{\textit{\underline{Citation}}: 
\textbf{Stevanovic, O., \& Wachter, J. (2026). Automating attack graph construction for agentic pentesting: Towards neuro-symbolic vulnerability hunting. In D. Hitaj et al. (Eds.), ESORICS 2026 workshops. Springer Nature Switzerland AG.
}}}

\author{Oliver Stevanovic
\\University of Udine
\\
via Palladio 833100 Udine, Italy \\
\texttt{oliverste@edu.aau.at} 
   \And
Jasmin Wachter \thanks{Corresponding author.}\\
University of Klagenfurt\\
Universitätsstraße 65-67, 9020 Klagenfurt, Austria\\
\texttt{jasmin.wachter@aau.at}
}

\begin{document}
\maketitle

\begin{abstract}
Logic attack graphs grounded in scanner output provide explicit and auditable attack path reasoning LLM-based agents lack. Integrating symbolic frameworks such as MulVAL to contemporary security workflows or agentic pipelines, however, requires translating scanner evidence to initial facts, and creating domain-specific rules. We present a semi-automated pipeline that addresses this interoperability problem and depict its feasibility in a web-security case study. Our pipeline parses findings from Trivy, Semgrep, and Nmap into MulVAL predicates and uses an LLM-assisted process to construct domain-specific Datalog rules linking scanner-detectable evidence to attack techniques. MulVAL/XSB then performs symbolic inference to generate structured attack paths. We evaluate the attack-graph construction infrastructure on 54 web Capture-the-Flag tasks from CyBench within an agentic pipeline (\texttt{Hybrid Reasoner}); we do not evaluate the performance of the downstream agent. Every task produced at least one goal-reaching graph, and we achieve mean ground-truth vulnerability coverage of 53.7\%, with 51.9\% achieving full coverage; mean noise-path rate was 83.9\%. With median end-to-end time of 24.9 s (MulVAL reasoning: 2.7 s) the pipeline is feasible and runtime-practical for agentic workflows, but predicate coverage, rule coverage, and path precision remain limiting factors. Next steps include semantic rule validation and agent-level comparison for graph-guided pentesting.

\end{abstract}

% keywords can be removed

\section{Introduction}
Agents based on Large Language Models (LLMs), in particular ReAct frameworks, are increasingly being investigated for automated penetration testing~\cite{MayoralVilchesetal2025,happe2025llms,mayoral-vilches2025cai}. Such systems can coordinate reconnaissance and exploitation tools and generate candidate actions from natural language context. However, their reasoning is commonly grounded in prompts, transcripts, or other unstructured states; decision making is \textit{ad hoc} and \textit{implicit}. This makes it difficult to audit the decision making process: to inspect which observed conditions support a proposed attack step, to reproduce multi-step reasoning, or to exchange structured security knowledge between scanners, reasoning engines, and agents.

\textit{Logic} attack graphs \cite{ou2005mulval} offer a complementary, symbolic representation. They are formal attack hypothesis models that systematically map vulnerabilities and attack steps in networks. These formalisms enable intuitive visualization and systemic analysis of security posture, supporting vulnerability prioritization and network hardening \cite{wachter2023graph}. Despite their utility, logic attack graphs remain underutilized in modern agentic security workflows: Grounded in declarative reasoning, they encode security-relevant system facts and inference rules explicitly, allowing a symbolic engine to derive attack paths with traceable dependencies \cite{ou2005mulval,ou2006scalable}. In principle, logic attack graphs could provide structured, explainable and verifiable context for a pen-testing agent, decoupling red-team planning from execution planning in agentic penetration testing. While initial evidence from \texttt{Incalmo} framework \cite{incalmo2026singer} suggests a performance benefit of decoupling high-level attack reasoning from attack execution, their analysis is LLM-based and does not involve any \textit{symbolic} attack reasoning extending their architecture for verifiable and auditable attack-path reasoning. 

The practical obstacle before the utility of symbolic attack graphs can be tested in LLM-agentic environments is the \textit{interoperability gap between LLM-based and logic-based attack-reasoning}: Established symbolic attack-graph systems such as MulVAL require an initial fact base describing the target and a rule base describing how observed conditions enable attack steps. MulVAL, however, was designed around particular vulnerability and configuration sources, while contemporary web-security workflows and agents typically use heterogeneous tools such as static analyzers, vulnerability scanners, and network scanners. Their findings do not directly instantiate MulVAL predicates, and generic MulVAL rules do not necessarily express domain-specific attack patterns. Connecting scanner output to a usable symbolic model therefore requires parsers, a shared predicate vocabulary, and domain-specific rules---work that is often manual and difficult to maintain~\cite{tayouri2023survey}. Moreover, no systematic approaches have been established to guide this integration work.

\paragraph{Bridging the Gap: Towards Neuro-Symbolic Vulnerability Hunting.} This paper investigates this \textbf{scanner-to-symbolic interoperability problem}. We present a semi-automated pipeline that (1) translates output from vulnerability and network scanners~\footnote{In our web case study we employ Trivy, Semgrep, and Nmap} into an initial MulVAL fact base; (2) uses an LLM-assisted process to construct domain-specific Datalog rules ~\cite{gallaire1984logic}; (3) connects low-level, scanner-detectable evidence to higher-level attack techniques through a two-layer predicate architecture; and (4) invokes MulVAL/XSB ~\cite{sagonas2023xsb} to derive structured attack paths. The generated paths can be supplied to a downstream agent, but the present study evaluates only the infrastructure that constructs them.

Our approach is \textit{neuro-symbolic} in two ways: architecturally, a neural language model assists in producing symbolic artifacts during offline knowledge-base construction, while MulVAL/XSB performs runtime logical inference. Downstream, the resulting attack graph provides a symbolic layer complementing the neural agentic pentesting framework. We stress, however, that while the present prototype is integrated in a downstream agent, the present study evaluates only the infrastructure that constructs them. It does not implement a neuro-symbolic feedback loop, nor does it establish that the neural component produces semantically correct rules without human review. Its contribution is therefore a step \textit{towards} \textit{neuro-symbolic} vulnerability hunting: a practical method to close the interoperability gap, customizing and populating an existing symbolic engine from contemporary security sources so that attack graph-guided agent experiments become possible.
\subsection{Research Question and Contribution}
To achieve our objective, we propose a semi-automatic neuro-symbolic pipeline that combines LLMs and Datalog for customized cybersecurity topology modeling. Specifically, we address the following \textbf{Research Question:} \textit{To what extent can an LLM-assisted scanner-to-MulVAL pipeline generate executable and vulnerability-relevant attack graphs for web-security tasks, and what limitations arise from the resulting predicates and rules?}

We ground our methodology in MITRE ATT\&CK framework as a semantic backbone and demonstrate the customization of our symbolic attack graph generation engine in a web-security case study. This case study builds on OWASP Top 10 web vulnerabilities and depicts a proof-of-concept of a real-world vulnerability hunting pipeline. 

In particular, we evaluate whether the generated custom attack graphs produce structurally valid and vulnerability-relevant paths for web security assessments. We evaluate the pipeline on a corpus of 54 web-oriented tasks drawn from the CyBench benchmark \cite{cybench2024} collection. For each task, we measure (1) \emph{goal reachability}---whether symbolic inference produces a goal-reaching graph (2) \emph{vulnerability fidelity}---whether generated paths contain the ground-truth vulnerability classes required by the task, (3) \emph{path noise}---how many paths are unrelated to those classes, and (4) \emph{generation time}---how long the pipeline requires. The pipeline produced a goal-reaching graph for all 54 tasks and completed in a median of 24.9 s, with a median MulVAL reasoning time of 2.7 s. However, mean vulnerability coverage was 53.7\%; only 51.9\% of tasks achieved full coverage, and the mean noise-path rate was 83.9\%. The results therefore show that automated construction is feasible and computationally practical, but the symbolic model is not yet sufficiently complete or precise to assume reliable downstream agent guidance.

This paper makes three contributions:
\begin{enumerate}
    \item A \textbf{scanner-to-MulVAL architecture} that separates offline knowledge-base construction from runtime graph generation and links scanner evidence to higher-level attack techniques through a two-layer predicate and rule structure.
\item A \textbf{proof-of-concept implementation for web security} that converts findings from multiple contemporary scanners into MulVAL facts and uses an LLM-assisted process to extend the domain rule base.
\item An \textbf{empirical evaluation on 54 CyBench web tasks} that reports graph reachability, vulnerability coverage, path noise, and runtime, thereby identifying concrete limitations in the current predicate mappings and rules.
\end{enumerate}
These contributions establish the infrastructure and methodology necessary for subsequent agent-performance studies; they do not establish that symbolic attack-graph guidance improves exploitation success, efficiency, safety, or reasoning quality. A controlled comparison among unguided agents, scanner-guided agents, and graph-guided agents is a separate next step left for future work.

\section{Related Work and Concepts}
\paragraph{Attack Graphs and MulVAL}Attack graphs (AG) are formal models that compose security relevant conditions, assets, and attacker actions into possible attack paths \cite{wachter2023graph}. In logic attack graph formalisms, nodes represent network states or vulnerabilities, while directed edges denote causal dependencies and exploit transitions. The \textit{MulVAL (Multihost, Multistage Vulnerability Analysis Language)} framework \cite{ou2005mulval,ou2006scalable} pioneered this approach using first-order logic and symbolic reasoning to automatically construct attack graphs. MulVAL is based on Datalog, combining an \textit{initial fact base} (predicates representing network configurations) with an \textit{exploit rule base} (inference rules for deducing new facts). The original framework converts vulnerability scanner output (OVAL format \cite{acm:oval2022}) and vulnerability data from the ICAT database \cite{MellICAT} into Datalog clauses. Network configurations, vulnerabilities, and hosts are encoded as predicates, while attacker behavior and privilege escalation are encoded as generic reasoning rules. The resulting attack graph is generated through logical deduction using the \texttt{XSB Prolog} engine with tabling for efficient polynomial-time scaling \cite{rao1997xsb}. 

Subsequent work has extended its scalability, representation, and attack-scenario coverage \cite{saha2008extending,tayouri2023survey,gandhi2026atag}. The continuing difficulty is customization. Contemporary tools used in agentic penetration-testing emit heterogeneous findings at different levels of abstraction, while a useful MulVAL model requires a stable predicate vocabulary and rules that connect those findings to attacker capabilities. Existing MulVAL extensions demonstrate the flexibility of logical attack graphs, but also show that adapting the fact and rule bases to new domains and data sources remains substantial engineering work \cite{tayouri2023survey}.  Our work addresses this interoperability layer and illustrates the methodology for web-security tasks by translating findings from Trivy, Semgrep, and Nmap into MulVAL facts and by connecting scanner-detectable evidence to higher-level attack techniques through domain-specific rules. We retain MulVAL/XSB as the inference engine and do not introduce a new attack-graph formalism or reasoning algorithm.

\paragraph{Language models constructing symbolic knowledge} Language models have been studied as interfaces for translating natural-language specifications into logic programs and other structured representations. Recent work demonstrates that smaller models paired with symbolic reasoning can match larger models alone \cite{alviano2025integrating}. Specifically, \cite{Coppolillo2026} addresses code generation correctness for Datalog/ASP at scale, while \cite{Alviano2025APE} evaluates open-source LLMs for semantic parsing into Datalog representations. These works establish baseline performance for Language-Model-to-Datalog translation tasks using custom and off-the-shelf LLMs~\footnote{In this work, we employed Claude Opus 4.5 and GPT-4o.}. Our approach leverages rules initially generated from AI using natural language input, studying whether resulting attack graphs produce structurally valid, vulnerability-relevant paths for web security assessments. Measuring rule generation accuracy for security-specific MulVAL rules is deferred to future work.

\paragraph{LLM-based pentesting and (symbolic) attack models}
Recent systems use language models to plan and execute penetration-testing activities, coordinate tools, and interact with challenge environments. In LLM-agentic ReAct frameworks, agent state and guidance are commonly conveyed through natural-language prompts, tool output, or retrieved text. Such LLM-only agentic pentesting tools operating without top-level abstract reasoning, have their shortcomings. Singer et al.~\cite{incalmo2026singer} performed an error analysis of agentic pentesting in long-horizon multi-host environments, showing that existing ReAct-style systems may pursue tasks irrelevant to the challenge, execute actions incorrectly, lose track of assets and bloat context. This impacts their effectiveness in multi-step reasoning and long-horizon challenges. Their framework \texttt{Incalmo} addresses this limitation by separating high-level attack planning from execution, using high-level LLM generated attack graphs and maintaining environment states. They empirically illustrate the benefit of explicit planning abstractions and structured state maintenance in agentic pipelines. \texttt{Incalmo}'s planning layer is, however, LLM-based and leaving the problem of the benefit of high-level \textit{symbolic} attack reasoning open, in particular the problem of construction logic-based attack graphs from heterogeneous vulnerability evidence inherently to agentic pentesting pipelines. 

Several systems combine LLMs with formal attack models for security analysis. Wang et al. ~\cite{wang2024sands} extract precondition–action–postcondition structures from cyber-threat-intelligence reports and compiles them into PDDL actions for attack planning. Unlike our approach, however, they employ cyber threat intelligence (CTI) reports to customize the planning objective and use planners, not MulVAL-style monotonic reasoning. Hou et al.'s recent preprint~\cite{Hou2026} also grounds their attack models in CTI reports, processing narrative reports to Datalog-style rules to extract reachable attack chains. Their framework is CTI-driven and grounded in attack stages, while our approach is hierarchical and tactic/technique-driven, allowing for domain customization at a lower hierarchical level. Other related work includes EntaiLLM,~\cite{Mukherji2026EntailLLM} which treats LLM-suggested attack paths as hypotheses and only allows for their execution if they are entailed by a domain-knowledge graph to analyze decompiled binaries. Related but different, the ATAG framework~\cite{gandhi2026atag} extends MulVAL logic based AG-tool to understand attacks on multi-agent AI systems and analyze the risks associated with AI-agent applications as the domain model. 

\paragraph{Positioning of this work}
Our work addresses a different evidence-to-symbolic interoperability problem: constructing customized MulVAL logic attack graphs instantiated from heterogeneous scanner evidence. We present a semi-automated pipeline that translates output from vulnerability and network scanners into an initial MulVAL fact base using a two-layer predicate and rule architecture. The pipeline translates findings from Trivy, Semgrep, and Nmap into an initial MulVAL fact base and connects low-level, scanner-detectable evidence to higher-level MITRE ATT\&CK techniques using connector predicates. On both levels, an LLM assists with offline rule construction, while runtime inference remains entirely within MulVAL/XSB. Unlike general language-to-logic studies, we evaluate the assembled security artifacts against task-specific vulnerability annotations, measuring goal reachability, vulnerability coverage, path noise, and runtime rather than the intrinsic accuracy of LLM-generated rules. The resulting graphs can be supplied to the \texttt{Hybrid Reasoner}, but the present study evaluates graph-construction feasibility rather than whether graph guidance improves agent performance. A rule-level test suite and expert validation as well as agent ablations are necessary follow-up work. Moreover, since automatic verifyers/validators are currently missing in the pipeline, we refer to the methodology as \textit{semi-automated};  manual inspection of web-relevant rules with respect to semantic correctness and deduplication checks with respect to generated predicates were performed.

\section{Framework Architecture}
\begin{figure}[h!]
    \centering
    \includegraphics[width=0.7\linewidth]{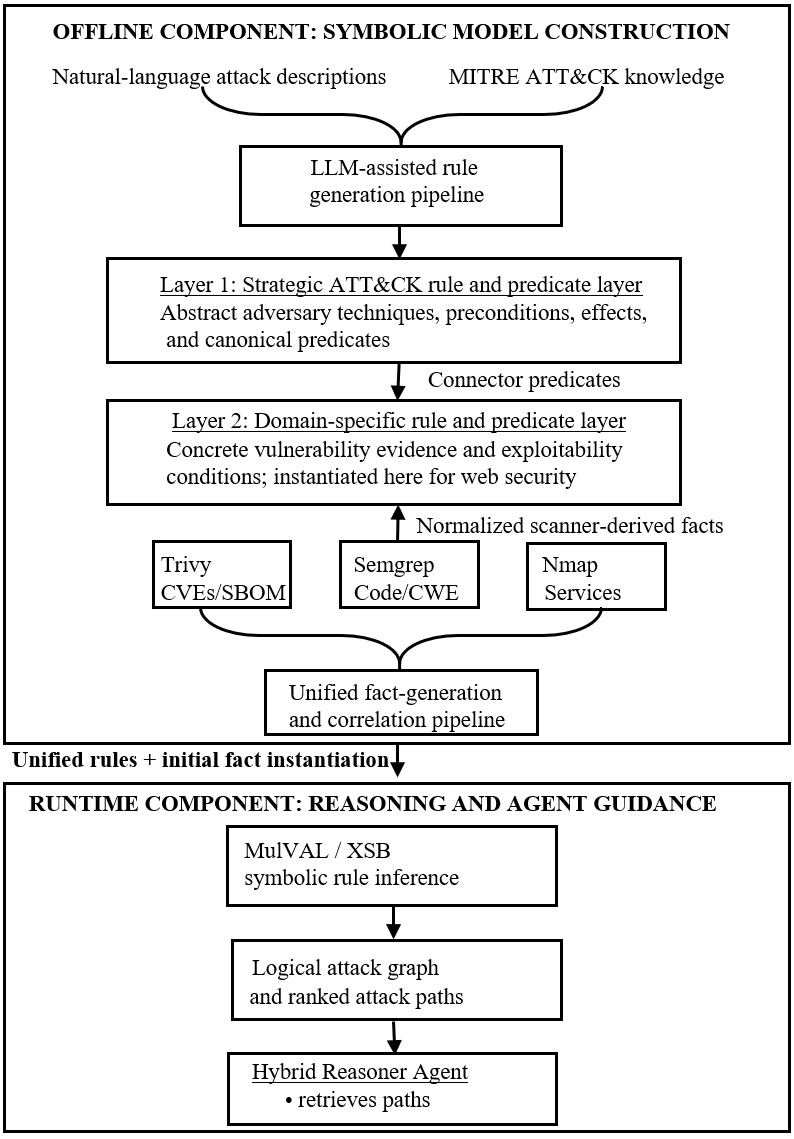}
    \caption{Framework architecture. The offline component constructs a hierarchical MulVAL knowledge base. A strategic MITRE ATT\&CK layer is connected to domain-specific rules through connector predicates, while scanner findings are correlated and translated into initial facts. At runtime, MulVAL/XSB derives logically justified attack paths to be provided as structured guidance to the \texttt{Hybrid Reasoner} agent as context.
}
    \label{fig:component}
\end{figure}

Decoupling abstract attack step reasoning from exploit execution was shown to enhance multi-step reasoning ~\cite{incalmo2026singer}; whether logic attack graphs provide the same benefit is an open problem. We propose an \textit{Exploit Rule and Initial Fact Generation} methodology to create the infrastructure necessary for such endeavors: a semi-automated pipeline to extend MulVAL with updated interaction rules for contemporary vulnerabilities and connect it to various scanner sources. We employ a two-component architecture: \textit{Offline Rule and Fact Generation} separate from \textit{Runtime Attack Graph Generation and Agentic Integration}, see Fig. \ref{fig:component}.\\

\noindent\textbf{Offline Component:} \textit{Rule and Fact Generation Pipeline.}
\begin{itemize}
\item[i)] \textit{Interaction Rules Generation} (Section \ref{sec:LLMgen}f): A semi-automated LLM-pipeline converts attack descriptions into Datalog rules, enforcing a two-level hierarchy-top-level MITRE ATT\&CK predicates ensure consistency, while customized second-level rules ground transitions in scanner-detectable evidence.~\footnote{Attack graph quality depends on the syntactical and semantic correctness of generated rules; formal validation is out of scope for this PoC.}
\item[ii)] \textit{Predicate Generation Pipeline (Section \ref{sec:rules2})}: An automated pipeline converts vulnerability scanner output into MulVAL facts, instantiating the initial fact base via custom parsers and connector predicates.
\end{itemize}

\vspace{0.15cm}
\noindent\textbf{Runtime Component:} \textit{Attack Graph Generation and Agentic Integration,} see (Section \ref{sec:graphconst})
\begin{itemize}
\item[i)] \textit{Graph Construction}: MulVAL generates attack graphs from the customized predicate and fact base.
\item[ii)] \textit{Agentic Integration}: CAI red-team agents utilize attack graphs as structured context, forming \texttt{Hybrid Reasoner} Agents.~\footnote{Agentic performance evaluation pending.}
\end{itemize}

\subsection{MulVAL Rule Generation and Customization Pipeline}
\label{sec:LLMgen}
At the highest level, we employed a semi-automated pipeline (see Section \ref{sec:rulesTop}) to create the MITRE ATT\&CK techniques top-level rule base. This provides a comprehensive semantic framework --- each technique encodes adversary behavior, preconditions, and postconditions that translate into logical transitions \cite{ou2005mulval,wachter2023graph,mitre_attack} within MulVAL's rule base. MITRE techniques are intentionally abstract and do not directly encode low-level exploit signals required by pen-testing agents.

To bridge this gap, we introduce a second, customizable layer which grounds abstract ATT\&CK transitions in scanner-detectable evidence --- see Section \ref{sec:rules2}. In our case study, this layer focuses on concrete web vulnerabilities (drawn from the OWASP Top 10 ~\cite{owasp}). Rather than treating these layers independently, we implement a connector predicate architecture that links lower-level findings --- such as endpoint exposure, weakness class, and attacker reachability --- to higher-level ATT\&CK-derived rules through staged derivations. This approach yields a comprehensive and maintainable ruleset that is large enough to capture realistic attack scenarios yet modular enough to be easily customized, extended, or adapted to organization-specific threat models without manual rewriting of hundreds of individual rules.

\subsection{Top-Level Rule Generation Pipeline (MITRE ATT\&CK)}\label{sec:rulesTop}
The \textit{Top-Level Rule Generation Pipeline} follows a semi-automated approach, involves three steps: (1) creating the \textit{Initial Predicate Vocabulary} established from MITRE ATT\&CK, which we reviewed in a Human-in-the-Loop manner. New rules generated in the LLM-pipeline must reuse this vocabulary; introduction of additional predicates is restricted to strictly necessary cases, preventing proliferation that would degrade interoperability and readability. In Step (2), \textit{Rule Generation}, the rules (Top-level techniques) are created using an LLM-pipeline.~\footnote{We employed Claude Opus 4.5 and GPT-4o for rule generation, while the methodology is agnostic of the underlying LLM.} Each generated rule corresponds to one technique and is materialized as a standalone \texttt{.rule} file with a machine-readable header and Datalog rule block. Rule bodies encode preconditions via scoped intermediate predicates feeding clauses with canonical or domain-specific derived predicate heads.
Finally, the \textit{Postprocessing} stage --- Step (3) --- merges rule files and predicate 
registries into a unified knowledge base through a layered construction: primitive and 
derived predicates are declared first, followed by memoization directives to prevent 
pathological derivation chains in cyclic graphs~\cite{rao1997xsb}.

\paragraph{Second-level Rule Customization} \label{sec:rules2}

The First-level rule and predicate layer provides a strategic backbone; techniques are intentionally abstract and do not directly encode the low-level exploit signals that a penetration-testing agent must follow. For (automatic) customization, we introduced a second expansion layer, designed as a bridge that links concrete findings to higher-level ATT\&CK transitions to facilitate future interoperability. Moreover, the second layer bridges network and vulnerability scanner output: scanner outputs become predicates through connector rules that ground abstract technique transitions in scanner-detectable evidence, facilitating future interoperability.

In the second layer, application-specific predicates derived from scanner evidence establish concrete exploitability conditions. Rather than manual authoring, this fact base is systematically synthesized from multiple scanners through the \texttt{UnifiedPipeline} (Section~\ref{sec:unified}). Intermediate predicates then map concrete evidence to ATT\&CK-compatible abstractions, bridging tactical exploit guidance to strategic technique-level reasoning. 
\subsection{Populating the Initial Facts --- A Unified Pipeline }
\label{sec:unified}

The \texttt{UnifiedPipeline} orchestrates second-level predicate instantiation through 
three phases: First, in \textit{scanner execution} we collect evidence: Second, \textit{evidence correlation} merges network and vulnerability scanner findings; Third, \textit{predicate synthesis} translates normalized evidence into MulVAL predicates, initializing the fact base. 

\paragraph{MulVAL Rule Customization: A Web Case Study}
In our web vulnerability case study (OWASP Top 10), the pipeline integrates Trivy for vulnerability and configuration scanning, and Semgrep for semantic code pattern analysis. 
Trivy generates Software Bill of Materials and vulnerability predicates; Semgrep identifies coding vulnerabilities mapped to CWE. Scanner outputs are correlated and normalized through a parser layer that transforms them into MulVAL-compatible predicates capturing host topology, 
exposed services, vulnerability evidence, and web context. These predicates activate corresponding interaction rules, enabling comprehensive modeling of infrastructure and application-level attack vectors.
\begin{table}[t]
\centering
\begin{tabular}{|p{2cm}|p{5cm}|p{5cm}|}
\hline
\textbf{Category} & \textbf{Predicates} & \textbf{Description} \\

\hline
Network Topology & \scriptsize{\texttt{attacker\_at/1}}$^*$, \scriptsize{\texttt{host/1}}$^N$, \scriptsize{\texttt{connects/3}}$^N$, \scriptsize{\texttt{exposed/2}}$^N$ & \scriptsize{Foundational network context and attacker position} \\
\hline
Service Infrastructure & \scriptsize{\texttt{service/5}}$^N$, \scriptsize{\texttt{os/2}}$^N$, \scriptsize{\texttt{role/2}}$^*$, \scriptsize{\texttt{auth\_service/3}}$^*$ & \scriptsize{Discovered services, operating systems, and their security-relevant roles} \\
\hline
Vulnerability Evidence & \scriptsize{\texttt{vuln/3}}$^T$, \scriptsize{\texttt{exploit\_possible/1}}$^T$, \scriptsize{\texttt{remote\_service\_vulnerable/4}}$^T$& \scriptsize{Generic CVEs and correlated service vulnerabilities} \\
\hline
Web Application Context & \scriptsize{\texttt{web\_service/3}}$^*$, \scriptsize{\texttt{has\_user\_input/3}}$^S$, \scriptsize{\texttt{has\_login\_form/2}}$^S$, \scriptsize{\texttt{uses\_database/3}}$^S$ & \scriptsize{Web-specific attack surfaces and application architecture} \\
\hline Data and Assets & \scriptsize{\texttt{has\_data/2}}$^*$, \scriptsize{\texttt{has\_database\_records/1}}$^*$ & \scriptsize{Sensitive data resources such as databases} \\
\hline
\end{tabular}
\caption{Second Level Predicates: Web Security Case Study. Predicates are parsed from T$^T$ = Trivy, $^S$ = Semgrep, $^N$ = Nmap, $^*$ = Other/Inferred, marked by the corresponding superscripts.}
\label{table:security_model}
\end{table}

\section{Attack Graph Generation and Agentic Integration}
\label{sec:graphconst}

MulVAL goal generation in our framework is contextual: generic goals like \texttt{compromise(Host, User)} are 
supplemented with service-specific goals (e.g., \texttt{initial\_access(Host, User, Method)}) when discovered. This focuses MulVAL reasoning on relevant attack paths rather than exhaustive reachability graphs. The generated predicates from our web security case study are presented in Table \ref{table:security_model}. It is organized into distinct categories to ensure logical structuring and maintainability of the attack graph. Implementation details and source code are available in the accompanying GitHub repository~\cite{githydra}.

The \textit{Runtime Component} of our framework constructs concrete attack scenarios --- attack graphs --- grounded in the symbolic model, see Section \ref{sec:graphconst}. The intended use of the infrastructure is described below.

\paragraph{Attack Graph Construction}
Attack graphs are generated via symbolic inference on MulVAL/XSB \cite{ou2005mulval,rao1997xsb}. The unified predicates and interaction rules are evaluated in a containerized environment, producing a derivation trace that is parsed into structured attack paths. The final output contains metadata (total paths, attack goals, target host) and individual paths with ordered steps tagged as either base facts or rule applications, providing explicit logical justification for each edge.

\paragraph{Graph Integration to Agentic AI}
In our implementation, we customized the \texttt{Red Team} agent from the CAI agent suite as the reference agent and give it instructions to use the information in the attack graph~\cite{saha2008extending,wachter2023graph}. The resulting \texttt{Hybrid Reasoner} Agent is  defined as an instance of the CAI \texttt{Agent} class.  While a full analysis of the \texttt{Hybrid Reasoner} is out of scope of this paper, the infrastructural components responsible for attack graph generation are subjected to empirical evaluation.

\section{Proof-of-Concept --- Evaluation Methodology}
\begin{table}[htbp]
\centering
\caption{Benchmark evaluation metrics}
\label{tab:benchmark_metrics}
\begin{tabular}{|p{3.5cm}|p{8.5cm}|}
\hline
\textbf{Metric} & \textbf{Definition} \\
\hline
Goal reachability & \scriptsize{A binary indicator: task marked reachable if \texttt{paths.json} contains at least one path whose goal term contains \texttt{compromise()}}. \\
\hline
Mean vulnerability coverage & \scriptsize{Average fraction of required vulnerability classes found in at least one path across all tasks, based on manual ground-truth annotations from \texttt{web\_ctf\_inventory.csv}.} \\
\hline
Full-coverage rate & \scriptsize{Fraction of tasks where every required vulnerability class appears in the generated paths.} \\
\hline
Mean noise path rate & \scriptsize{Average fraction of generated paths that match none of the required class signatures and are not useful for guiding exploitation.} \\
\hline
Pipeline time & \scriptsize{Total generation time per task, excluding Docker container startup overhead.} \\
\hline
Scanner/predicate time & \scriptsize{Time spent on vulnerability scanning \& predicate generation.} \\
\hline
MulVAL reasoning time & \scriptsize{Time spent on attack graph reasoning \& path generation.} \\
\hline
\end{tabular}
\end{table}
To assess whether our framework produces structurally valid and vulnerability-relevant paths for web security assessments, we run the full symbolic pipeline over a corpus of web-oriented CTF (Capture The Flag) tasks drawn from the CyBench benchmark collection \cite{cybench2024}. The benchmark corpus is assembled from a static inventory CSV (\texttt{web\_ctf\_inventory.csv}) that references web challenges from two CyBench source collections: \texttt{hkcert-ctf} and \texttt{project-sekai-ctf}. Each entry carries a \texttt{required\_vulns} field with one or more ground-truth vulnerability classes (e.g.\ \texttt{xss}, \texttt{auth\_bypass}, \texttt{ssti}) assigned by manual inspection of each challenge's source code and published solutions. A task-discovery pass resolves each entry against the CyBench directory tree. The resulting corpus contains 54 web tasks spanning static and live deployment modes, all with ground-truth annotations.

\paragraph{Evaluation Setup}
For each task the benchmark runner executes the following sequential stages, timing each independently:
 \textit{(1) Scanner and predicate generation.} Static analyzers (Trivy, Semgrep, Nmap) generate predicates from the task environment; \textit{(2) MulVAL reasoning.} Predicates and rules are evaluated via symbolic inference, producing a derivation trace.
 \textit{(3) Trace parsing.} The derivation trace is parsed into structured attack paths and output artifacts. The evaluation then measures three properties: whether the pipeline reaches an objective state at all (\emph{goal reachability}), how faithfully the paths represent the vulnerability classes required by each challenge (\emph{vulnerability fidelity}, \emph{noise path rate}), and the wall-clock cost of each pipeline execution (\emph{generation time}) --- see Table \ref{tab:benchmark_metrics}. 

\section{Results and Discussion}
 \textit{Goal reachability} is satisfied for every task in the corpus: this means that for every challenge a graph is successfully created. This metric suggests that the rules produced at least one executable goal derivation per task. 
 
Regarding \textit{Generation Time} the scanner and predicate generation stage accounts for the dominant share of wall-clock time: the median MulVAL reasoning time is 2.7\,s, while total pipeline time has a median of 24.9\,s with a 90th percentile reaching 73.9\,s. The bottleneck is input predicate generation: information retrieval leveraging static scanners is time-consuming; Nmap in particular performs extensive verifications on the systems's open ports to identify sofware versions used in the challenges.

\begin{figure}[h!]
\centering
\includegraphics[
  angle=-90,
  origin=c,
  height=0.62\textheight
]{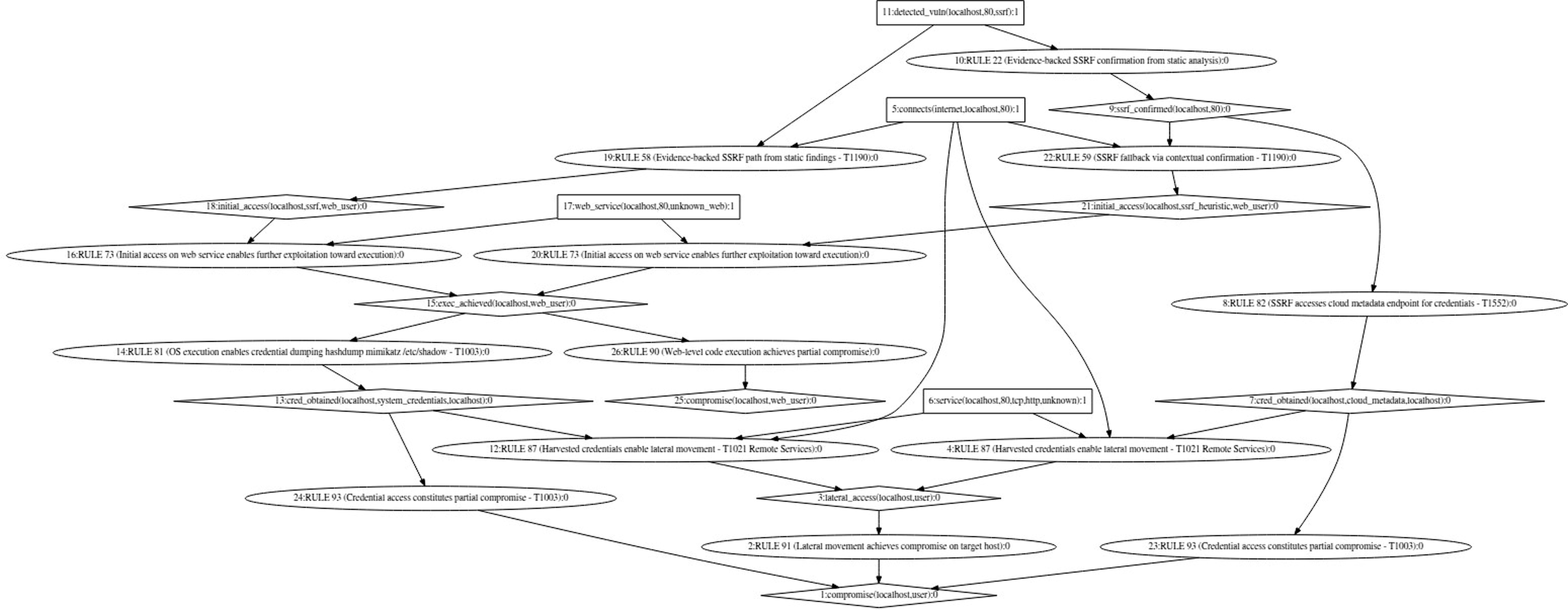}
\caption{An example logical attack graph from the \texttt{hkcert-ctf} challenge.}
\label{fig:nature}
\end{figure}

\textit{Vulnerability coverage} is evaluated against ground-truth annotations for all 54 tasks. Of these, 28 tasks (51.9\,\%) achieve full coverage (all required vulnerability classes present in at least one path), 2 tasks achieve partial coverage, and 24 tasks (44.4\,\%) produce zero coverage, meaning the generated paths contain no signature matching any ground-truth vulnerability that should be exploited to correctly gain the flag. The most frequently missing classes are \texttt{xss} (6 tasks), \texttt{auth\_bypass} (4 tasks), and \texttt{path\_traversal} (3 tasks), reflecting areas where the current interaction rules or Semgrep-to-predicate mappings do not yet produce enough evidence to integrate in the paths. The mean noise path percentage of 83.9\,\% indicates that a big portion of the generated paths refer to unintended exploits when it comes to the specific challenges from the benchmark. \\
At first glance this statistic could look extremely negative, because for the given benchmark, the attack graphs are bloated with unnecessary information the agent would have to prune while performing the assessment. While this issue should be addressed in future work, we stress that this metric is relative to the ground truth, i.e. the benchmarks annotated vulnerability classes. Thus, we treat noise paths as irrelevant to the ground truth, but stress that restricting the generation of attack hypothesis or the rule set to ground truth could create overfitting and have negative performance effects, preventing the agents from exploring alternative successful attack attempts. 
\begin{table}[htbp]
\centering
\caption{Aggregate benchmark results (54 web CTF tasks).}
\label{tab:benchmark_aggregate}
\begin{tabular}{|lr|lr|}
\hline
\textbf{Metric} & \textbf{Value} & \textbf{Metric} & \textbf{Value} \\
\hline
Tasks evaluated & 54 & Goal reachability rate & 100.0\,\% \\
\hline
Tasks \scriptsize{(w. ground-truth vul'n annotations)} & 54 / 54 & Mean vul'n coverage & 53.7\,\% \\
\hline
Full-coverage rate \scriptsize{(all vuln's present)} & 51.9\,\% & Mean noise path rate & 83.9\,\% \\
\hline
Pipeline time \scriptsize{(median)} & 24.9\,s & Pipeline time \scriptsize{(p90)} & 73.9\,s \\
\hline
Scanner/predicate time \scriptsize{(median)} & 12.4\,s & MulVAL reasoning time \scriptsize{(median)} & 2.7\,s \\
\hline
\end{tabular}
\end{table}
Table~\ref{tab:benchmark_aggregate} reports aggregate statistics over the full 54-task corpus while Table~\ref{tab:benchmark_vuln_multi} details results for the four \textit{multi-vulnerability tasks} that require more than one vulnerability class to solve, representing the hardest cases in the corpus.
\begin{table}[htbp]
\centering
\small
\caption{Benchmark results for multi-vulnerability tasks}
\label{tab:benchmark_vuln_multi}
\begin{tabular}{|p{4.3cm}|p{4.8cm}|r|r|}
\hline
\textbf{Task} & \textbf{Required classes} & \textbf{Coverage} & \textbf{All vul's} \\
\hline
\scriptsize{CTF-2022/42-prototype-sanityxss} & \scriptsize{prototype\_pollution; xss}             & 50.0\,\% & No  \\
\hline
\scriptsize{CTF-2023/15-wishlist}            & \scriptsize{xss; ssrf}                             & 100.0\,\% & Yes \\
\hline
\scriptsize{sekaictf-2023/chunky}            & \scriptsize{request\_smuggling; cache\_poisoning}  & 0.0\,\%  & No  \\
\hline
\scriptsize{sekaictf-2023/frog-waf}          & \scriptsize{ssti; cmdi}                            & 50.0\,\%  & No  \\
\hline
\end{tabular}
\end{table}
Only one of the four multi-vulnerability tasks achieves full coverage (CTF-2023/15-wishlist, requiring XSS and SSRF). The \texttt{chunky} task's required classes (\texttt{request\_smuggling} and \texttt{cache\_poisoning}) represent advanced HTTP-level attack patterns that the current rule base does not model with sufficient specificity. Future work could address this problem by introducing a wider range of vulnerabilities from the web domain, similarly to how MITRE techniques have been processed, to improve these rates. The remaining two tasks achieve partial coverage, where one of the two required classes is represented but the second is absent from all generated paths.

\paragraph{Discussion} The evaluation establishes that the pipeline can consistently produce executable derivations at modest runtime, while the lower vulnerability coverage and high path-noise rate show that the current attack graphs alone are not yet reliable enough for unqualified downstream agentic guidance. Nonetheless, operational feasibility is established which makes grounding LLM-agents in symbolic attack graph model technically plausible, offering potential in several application domains. First, similar to knowledge graphs in EntailLLM~\cite{Mukherji2026EntailLLM}, structured attack graphs could be studied as \textit{guardrails} for LLM-based pentesting agents, which could restrict exploitation to attack paths derived from encoded facts and rules and potentially reduce disruptive actions. Moreover, since every recommended action is backed by an explicit symbolic attack graph and derived from rules, the resulting attack paths become symbolically traceable and auditable at the derivation level. Lastly, the approach is particularly attractive for operational technology (OT) environments, where active penetration testing is often restricted by safety and availability requirements. In such settings, attack graphs could provide operators with structured attack hypotheses for planning, prioritizing assessments \textit{before} limited maintenance windows. Beyond agentic applications and attack orchestration, the same infrastructure could support traditional security engineering tasks such as network hardening~\cite{wachter2023patching,rass2023game}.

\section{Conclusion}
This work presents an initial investigation into neuro-symbolic vulnerability hunting by integrating symbolic AI methodologies within agentic pentesting workflows. The automated expansion of the MulVAL knowledge base to incorporate domain-specific attack patterns represents a foundational contribution toward systematic vulnerability discovery. A proof-of-concept case study on Web Security establishes a baseline for subsequent research and identifies critical dimensions for enhancement. The results in particular distinguish \textit{goal reachability} from \textit{vulnerability fidelity}. MulVAL reached a compromise goal for every task, showing that the generated facts and rules formed executable derivations. However, the substantially lower vulnerability coverage and high path-noise rate show that reachability alone is an insufficient quality criterion. Errors can arise because scanners fail to observe relevant evidence, parser mappings omit or misclassify predicates, rules do not cover the required vulnerability class, or rules admit paths that are logically executable but irrelevant to the solution.

\paragraph{Future Work and Limitations}
 As discussed in previous sections the pipeline's effectiveness is heavily dependent on the quality of the attack graph generation rules and input predicates. The present aggregate evaluation localizes the limitation to the combined pipeline but does not yet attribute error to individual rules or components. A component-level activation and provenance analysis is therefore required before the graphs can be treated as reliable guidance. 
 
Another open question relates to how structured attack graphs impact agentic pentesting performance beyond providing vulnerability hints. In particular, this involves comparing the (baseline) \texttt{Red Team Agent} and the \texttt{Hybrid Reasoner} in an explorative ablation study. Besides targeting prompt sensitivity effects, ablation should compare agentic performance across at least three conditions: \textit{(1) no graph input}, \textit{(2) raw scanner output}, and \textit{(3) structured attack graphs}, measuring relevant metrics, such as pass@3, token efficiency, tool calls, dead-end actions, success per scanner finding, and adherence to exploitable paths. Thus, directions for future work include: \textit{(i) Improving Input Predicate Quality.} Current scanner output is noisy and incomplete; more accurate static analyzers and verification mechanisms could enhance predicate quality. \textit{(ii) Expanding Rule Coverage.} The existing ruleset covers only a limited subset of vulnerabilities. Extending coverage to broader attack patterns is essential for more precise attack graphs. \textit{(iii) Provenance of Failure.} Verification and validation loops as well as a pipeline to attribute root cause of failure to individual rules or components should be developed. \textit{(iv) Leveraging Small Language Models for Rule Generation/Validation.} Scaling rule generation to diverse organizational contexts requires evaluation of SLM capabilities for security-specific tasks. A promising approach employs SLMs generate rule candidates in-context in privacy-constrained environments, while LLMs handle verification and syntactic checking via tool invocation. \textit{(v) Standardized Tool Integration.} Current validation loops lack interoperability across agentic frameworks. Standardized interfaces such as MCP-Solver~\cite{szeider2025mcpsolver} offer unified mechanisms for syntactical verification of LLM-generated rules.

% **********************************************
% * BIBLIOGRAPHY SECTION
% **********************************************
\bibliographystyle{plain} % Options: plain, alpha, unsrt, etc.
\bibliography{references} % This looks for a file named references.bib

@inproceedings{ou2006scalable,
  author    = {Ou, Xinming and others},
  title     = {A Scalable Approach to Attack Graph Generation},
  booktitle = {Proc. ACM CCS},
  pages     = {336--345},
  year      = {2006}
}

@article{gallaire1984logic,
  author  = {Gallaire, Herv{\'e} and others},
  title   = {Logic and Databases: A Deductive Approach},
  journal = {ACM Comput. Surv.},
  volume  = {16},
  number  = {2},
  pages   = {153--185},
  year    = {1984}
}

@article{MayoralVilchesetal2025,
  author  = {Mayoral-Vilches, V{\'i}ctor and Wachter, Jasmin and others},
  title   = {{CAI} Fluency: A Framework for Cybersecurity {AI} Fluency},
  journal = {arXiv:2508.13588},
  year    = {2025},
  doi     = {10.48550/arXiv.2508.13588}
}

@article{wachter2023graph,
  author  = {Wachter, Jasmin},
  title   = {Graph Models for Cybersecurity: A Survey},
  journal = {arXiv:2311.10050},
  year    = {2023},
  doi     = {10.48550/arXiv.2311.10050}
}

@article{Mukherji2026EntailLLM,  
author = {Mukherji, Kaustuv and others},  
title = {EntailLLM: Verifying LLM-Generated Vulnerability Discovery Paths with Domain Knowledge via Logic Programming},  
journal = {arXiv:2608.01763},  
year = {2026},  
eprint = {2608.01763},  
archivePrefix = {arXiv},  
primaryClass = {cs.CR}}

@article{incalmo2026singer,
author = {Singer, Brian and others},
booktitle = {2026 IEEE S\&P },
title = {{Incalmo: an Autonomous Llm-Assisted System for Red Teaming Multi-Host Networks}},
year = {2026},
volume = {},
ISSN = {},
pages = {4282-4300},
doi = {10.1109/SP63933.2026.00132},
publisher = {IEEE Computer Society},
address = {Los Alamitos, CA, USA},
month =May}

@inproceedings{saha2008extending,
  author    = {Saha, Diptikalyan},
  title     = {Extending Logical Attack Graphs for Efficient Vulnerability Analysis},
  booktitle = {Proc. ACM CCS},
  pages     = {63--74},
  year      = {2008}
}

@inproceedings{rao1997xsb,
  author    = {Rao, Prasad and others},
  title     = {{XSB}: A System for Efficiently Computing Well-Founded Semantics},
  booktitle = {Proc. LPNMR},
  pages     = {2--17},
  year      = {1997},
  publisher = {Springer}
}

@misc{MellICAT,
  author = {Mell, Peter},
  title  = {Identifying Critical Patches with {ICAT}},
  year   = {2000},
  note   = {NIST ITL Bulletin}
}

@misc{acm:oval2022,
  author = {{MITRE Corporation}},
  title  = {Open Vulnerability and Assessment Language ({OVAL})},
  url    = {https://oval.mitre.org/}
}

@article{cybench2024,
  author  = {Zhang, Andy Y. and others},
  title   = {Cybench: A Framework for Evaluating Cybersecurity Capabilities and Risk},
  journal = {arXiv:2408.08926},
  year    = {2024},
  doi     = {10.48550/arXiv.2408.08926}
}

@misc{owasp,
  author = {{OWASP Foundation}},
  title  = {{OWASP} Top 10: The Ten Most Critical Web Application Security Risks},
  year   = {2021},
  url    = {https://owasp.org/www-project-top-ten/},
  note   = {Accessed: 2026-03-17}
}

@misc{mitre_attack,
  author = {{MITRE ATT\&CK}},
  title  = {{ATT\&CK} Techniques},
  year   = {2026},
  url    = {https://attack.mitre.org/techniques/},
  note   = {Accessed: 2026-03-17}
}

@article{Hou2026,
  author = {Hou, Wenbo and others},
  title = {An Automated Framework for Extracting Reachable Attack Chains from Cyber Threat Intelligence Reports},
  journal = {arXiv:2607.19742},
  year = {2026},
  eprint = {2607.19742},
  archivePrefix = {arXiv},
  primaryClass = {cs.CR}
}

@inproceedings{wachter2023patching,
  title={Patching Security Vulnerabilities Using Stackelberg Security Games on Attack Graphs},
  author={Wachter, Jasmin},
  booktitle={FAIEMA 2023},
  pages={83--98},
  year={2023},
  organization={Springer}
}

@article{rass2023game,
  title={Game-theoretic APT defense: An experimental study on robotics},
  author={Rass, Stefan and others},
  journal={Computers \& Security},
  volume={132},
  pages={103328},
  year={2023},
  publisher={Elsevier}
}

@inproceedings{ou2005mulval,
  author    = {Ou, Xinming and others},
  title     = {{MulVAL}: A Logic-Based Network Security Analyzer},
  booktitle = {Proc. USENIX Security},
  pages     = {113--128},
  year      = {2005}
}

@inproceedings{Alviano2025APE,
  author    = {Alviano, Mario and others},
  title     = {A Preliminary Evaluation of Open-Source {LLMs} for Datalog-Based Semantic Parsing in the {ASVIN} Project},
  booktitle = {ICLP Workshops},
  year      = {2025}
}

@article{szeider2025mcpsolver,
  author  = {Szeider, Stefan},
  title   = {{MCP}-Solver: Integrating Language Models with Constraint Programming Systems},
  journal = {arXiv:2501.00539},
  year    = {2025},
  doi     = {10.48550/arXiv.2501.00539}
}

@inproceedings{alviano2025integrating,
  author    = {Alviano, Mario and others},
  title     = {Integrating Answer Set Programming and Large Language Models for Enhanced Structured Representation of Complex Knowledge in Natural Language},
  booktitle = {Proc. IJCAI},
  pages     = {4330--4338},
  year      = {2025}
}

@article{mayoral-vilches2025cai,
  author  = {Mayoral-Vilches, V{\'i}ctor and others},
  title   = {{CAI}: An Open, Bug Bounty-Ready Cybersecurity {AI}},
  journal = {arXiv:2504.06017},
  year    = {2025},
  doi     = {10.48550/arXiv.2504.06017}
}

@article{happe2025llms,
  author  = {Happe, Andreas and Cito, J{\"u}rgen},
  title   = {On the Surprising Efficacy of {LLMs} for Penetration Testing},
  journal = {arXiv:2507.00829},
  year    = {2025},
  doi     = {10.48550/arXiv.2507.00829}}

@article{tayouri2023survey,
  author  = {Tayouri, Dawit and others},
  title   = {A Survey of {MulVAL} Extensions and Their Attack-Scenario Coverage},
  journal = {IEEE Access},
  volume  = {11},
  pages   = {27974--27991},
  year    = {2023},
  doi     = {10.1109/ACCESS.2023.3248917}
}

@software{semgrep,
  author = {{r2c}},
  title  = {Semgrep},
  year   = {2019},
  note   = {Open-source static analysis tool}
}

@misc{sagonas2023xsb,
  author = {Sagonas, Konstantinos and others},
  title  = {{XSB}: The Thirty-Year Perspective on Tabled Prolog},
  year   = {2023},
  note   = {Open-source logic programming and database system}
}

@article{Coppolillo2026,
  author  = {Coppolillo, Erica and others},
  title   = {Fine-Tuning {LLMs} for Answer Set Programming},
  journal = {J. Intell. Inf. Syst.},
  volume  = {64},
  pages   = {653--685},
  year    = {2026},
  doi     = {10.1007/s10844-025-01017-4}
}

@article{wang2024sands, 
title={From sands to mansions: Towards automated cyberattack emulation with classical planning and large language models}, 
author={Wang, Lingzhi and others}, 
journal={arXiv:2407.16928}, 
year={2024} }

@inproceedings{gandhi2026atag,
  title={Atag: Ai-agent application threat assessment with attack graphs},
  author={Gandhi, Parth Atulbhai and others},
  booktitle={Proceedings of the ACM Asia CCS},
  pages={805--819},
  year={2026}
}

@misc{githydra,
author={Wachter, Jasmin and others},
title = {{HYbrid ReAsoner (Hydra)}},
year = {2026},
url = {https://github.com/JasminWachter/Hydra/},
note   = {https://github.com/JasminWachter/Hydra/}
}

\section*{Acknowledgments}
\subsection*{Use of AI}
LLMs were instrumental in generating domain-specific MulVAL rules from
natural-language attack descriptions, including Claude Opus 4.5 and
GPT-4o. Coding agents assisted with software development, while LLMs
supported manuscript preparation. All outputs were subject to manual
review by the authors before inclusion to the manuscript.
\subsection*{CRediT Author Contributions} \textit{O. Stevanovic:} Investigation, Validation, Methodology, Software, Writing–original draft; review \& editing. \textit{J. Wachter:} Conceptualization, Methodology, Supervision, Software, Writing–original draft; review \& editing
\subsection*{Competing Interests}
The authors have no competing interests to declare that are relevant
to the content of this article.

%Bibliography

\end{document}